\documentclass[10pt,a4paper]{article}
\usepackage{a4wide}

\usepackage[numbers, square, comma, sort&compress]{natbib}
\usepackage{amsmath}
\usepackage{amsfonts}
\usepackage{amssymb}
\usepackage[bookmarks = false, pdfstartview = FitV, colorlinks = false, linkcolor=green, urlcolor = blue]{hyperref}

\newcommand\bbm{\begin{bmatrix}}
\newcommand\ebm{\end{bmatrix}}
\newcommand\bpm{\begin{pmatrix}}
\newcommand\epm{\end{pmatrix}}
\newcommand\be{\begin{equation}}
\newcommand\ee{\end{equation}}
\newcommand{\half}{\frac12}
\newcommand{\eps}{\varepsilon}
\newcommand{\ket}[1]{\left\vert{#1}\right\rangle}
\newcommand{\bra}[1]{\left\langle{#1}\right\vert}
\newcommand{\kb}[2]{\vert#1\rangle\!\langle#2\vert}
\newcommand{\bk}[2]{\left\langle#1\vert#2\right\rangle}
\newcommand{\di}{\mathrm{d}}

\newcommand{\m}[1]{\mathcal{#1}}

\newcommand{\ii}{\mathrm{i}}
\newcommand{\Ee}{\mathrm{e}}

\newcommand\RF{\text{\tiny{RF}}}
\newcommand\SG{\text{\tiny{SG}}}
\DeclareMathOperator{\Tr}{Tr}

\begin{document}
\bibliographystyle{ieeetr}

\title{WKB analysis of relativistic Stern-Gerlach measurements}
\author{\\\\
Matthew C. Palmer$^1$, Maki Takahashi$^1$, and Hans F. Westman$^{1,2,3}$\\
{\small \it $^{1}$School of Physics, The University of Sydney, Sydney, NSW 2006, Australia}\\%
{\small \it $^{2}$Centre for Time, The University of Sydney, Sydney, NSW 2006, Australia}\\%
{\small \it $^{3}$Perimeter Institute for Theoretical Physics, Waterloo, Ontario N2L 2Y5, Canada}\\[2mm]%
}

\date{\today}

\maketitle

\begin{abstract}

Spin is an important quantum degree of freedom in relativistic quantum information theory. This paper provides a first-principles derivation of the observable corresponding to a Stern-Gerlach measurement with relativistic particle velocity. The specific mathematical form of the Stern-Gerlach operator is established using the transformation properties of the electromagnetic field. To confirm that this is indeed the correct operator we provide a detailed analysis of the Stern-Gerlach measurement process. We do this by applying a WKB approximation to the minimally coupled Dirac equation describing an interaction between a massive fermion and an electromagnetic field. Making use of the superposition principle we show that the $+1$ and $-1$ spin eigenstates of the proposed spin operator are split into separate packets due to the inhomogeneity of the Stern-Gerlach magnetic field. The operator we obtain is dependent on the momentum between particle and Stern-Gerlach apparatus, and is mathematically distinct from two other commonly used operators. The consequences for quantum tomography are considered.

\end{abstract}


\section{Introduction}

Over the past decade or so there has been a growing interest into the field of relativistic quantum information \cite{PTW11, Pablo1111,Czachor,Ternotworol,PeresTerno04,PeresScudoTerno02,BartlettTerno05, Caban1,Caban2, Friis-relent, Landulfo,Alsing,TerashimaUeda03,Louko, Fuentes,Brodutch11,BDT11}, the goal being to develop a mathematical framework which combines aspects of both relativity and quantum information theory. The aim of this program is primarily to shed light on the relationship between these two cornerstones of physics but also to  investigate possible near future applications in areas such as long range quantum communication in which relativistic effects cannot be neglected \cite{PTW11}.

One of the necessary features for this program is a relativistic measurement formalism, i.e.\;a recipe for extracting empirical predictions given a measurement setup and a quantum state. First and foremost this formalism is required to be Lorentz invariant in the sense that the predicted statistics should be independent of the reference frame in which we choose to describe the experiment. In order to do this it will be convenient to introduce a notation which is manifestly Lorentz covariant. As result of this we will be required to not only recast Hermitian observables into this relativistic notation but also replace the standard non-relativistic inner product. The advantage of doing this will not only be to develop a relativistic measurement formalism which is manifestly Lorentz invariant but, as we shall see, will also greatly simplify the derivation of a relativistic Stern-Gerlach measurement operator.

This paper will be concerned with relativistic spin measurements. In the literature there have been several proposals dating back to the 1960's for relativistic spin operators and these have been studied in the context of quantum field theory for various reasons (see e.g. \cite{FoldyWouthuysen,HehlNi,Ryder98,Ryder99,Mashhoon95}). More recently, these operators have been used in relativistic quantum information theory to predict measurement statistics for relativistic spin measurements \cite{Czachor,Ternotworol,Caban1, Caban2,Friis-relent}. The approach of this paper will not follow these proposals. Rather we will follow a strictly operational approach, where we will expand on results developed in \cite{PTW11, Pablo1111}. Specifically, we will derive the relevant spin operator for a Stern-Gerlach measurement of a relativistic massive fermion. Importantly, our operational approach yields a Hermitian spin operator which is mathematically distinct from these previous proposals.

The outline of this paper is as follows: We will begin by reviewing the manifestly Lorentz covariant formalism developed in \cite{PTW11}. We will then derive the relativistic Stern-Gerlach spin observable by modelling a  Stern-Gerlach measurement. Firstly, the specific mathematical form of the Stern-Gerlach operator is established using the transformation properties of the electromagnetic field. Next, to confirm that this is indeed the correct operator we provide a detailed analysis of the Stern-Gerlach measurement process. We do this by applying a WKB approximation to the minimally coupled Dirac equation describing an interaction between a massive fermion and an electromagnetic field. Making use of the superposition principle we show that the $+1$ and $-1$ spin eigenstates of the proposed spin operator are split into separate packets due to the inhomogeneity of the Stern-Gerlach magnetic field. We conclude by discussing the consequences for quantum tomography.

\section{Mathematical description of spin qubits \label{sec-mathdesc}}

Before we can describe a quantum mechanical relativistic spin measurement, we must first specify how one can represent spin in such relativistic scenarios, and furthermore specify what the transformation properties under the Lorentz group are for such a representation. Having identified a particular representation we will then need to develop a measurement formalism and identify the form of Hermitian observables.

There are two main ways in which one can represent spin. A common way is to make use of the Wigner representations \cite{Weinberg}, which are in fact infinite dimensional unitary representations of the Poincar\'e group.  This group has the added symmetry of translational invariance, and consequently Wigner basis states $\ket{p,\sigma}$ are labelled with both momentum $p$ and spin $\sigma=1,2$. In this representation the momentum $p$ transforms under the Lorentz group according to $p^\alpha\to \Lambda^\alpha_{\ \beta} p^\beta$, where $\Lambda$ is a general Lorentz transformation and $\alpha, \beta,\ldots = 0,1,2,3$ are spacetime indices, with spacetime metric $\eta^{\alpha\beta} = \mathrm{diag}(1,-1,-1,-1)$. However, the spin component strictly transforms under the {\it Wigner rotations} and constitutes a representation of what is called {\it Wigner's little group}, which is isomorphic to $SU(2)$. Specifically, Wigner's little group consists of the set of Lorentz transformations under which the `standard' momentum $p^{\alpha} = m\delta^\alpha_0$ of a particle with mass $m$ is left invariant \cite{Weinberg}, where $\delta^\alpha_{\beta}$ is the Kronecker delta symbol.

The representation used in this paper, and in \cite{PTW11} to which we refer the reader for further details and theoretical background, is distinct from the Wigner representation and deemphasises the use of the Wigner rotations. Here the mathematical object representing spin is an $SL(2,\mathbb C)$ spinor. This is a two-component complex-valued object $\psi_A$ with index $A=1,2$ that transforms covariantly under the spin-$\half$ representation of the Lorentz group, i.e.\;by $\psi_A\to\Lambda_A^{\ B}\psi_B$. This transformation law differs from the Wigner rotations, which are spatial rotations defined using a preferred frame. The reason why we adopt this alternative but equivalent representation of spin is primarily because we will insist on manifest Lorentz covariance which will greatly simplify our derivation of our relativistic spin operator. However, the results of this paper are presented in a form in which they can be interpreted in the Wigner representation if desired.

As this formalism may not be familiar to some, we will first briefly summarize the notation and the key features. Specifically, we will review spinor notation, the definition of an inner product, the modified notion of unitarity, and finally Hermitian operators and observables.

\subsection{Representation of Spin and spinor notation}

A spinor is fundamentally a two-component complex vector $\psi_{A}$ living in a two-dimensional complex vector space $W$. Here spinors are taken to be irreducible representations of $SL(2,\mathbb{C})$, which forms the double cover of the identity component of the Lorentz group, $SO^{+}(1,3)$. As such, $W$ carries a spin-$\half$ representation of the Lorentz group. In the spirit of Relativity we will use a geometric notation similar to that used for tensors. Therefore, a spinor $\psi_{A}$ carries an $SL(2,\mathbb C)$ spinor index $A=1,2$.

Complex conjugation takes a spinor $\psi_{A}\in W$ to a spinor $\overline{\psi_{A}} = \overline{\psi}_{A'}\in\overline{W}$, the conjugate space of $W$. We distinguish the elements of $\overline W$ by placing a prime on the spinor index: $A'$ (as is the notation commonly used in treatments of spinors \cite{Wald,Bailin,Penrose}). The summation of indices then follows the Einstein summation convention: We can only contract when one index appears as superscript and the other as subscript, and only when the indices are either both primed or both unprimed, e.g. $\phi_A\psi^A$ and $\xi_{A'} \chi^{A'}$, but not $\bar\phi_{A'} \psi^{A}$.

\subsection{Lorentz group and $SL(2,\mathbb C)$}

 We are concerned with spin of a massive fermion such as an electron. Such an object is usually taken to be represented by a four-component Dirac field $\Psi(x)$, which constitutes a reducible spin-$\half$ representation of the Lorentz group. At the same time we would like a qubit representation of our spin-$\half$ system, so it is natural to use a two-dimensional object. Such a representation can be found by working with the Dirac field in the Weyl representation. In this representation the Dirac field splits into two 2-component  $SL(2,\mathbb C)$ spinors $\Psi(x) = (\phi_A(x),\chi^{A'}(x))$ which constitute the left and right handed irreducible spin-$\half$ representations of the Lorentz group. We will represent qubits with the two-component left-handed Weyl spinor field $\phi_{A}(x)$. Working instead with the right-handed component $\chi^{A'}$ would yield the same results.

\subsubsection{$SL(2,\mathbb C)$}

We now turn to the Lorentz group. In the Weyl representation, the Dirac gamma matrices take on the form
\[
\gamma^{\alpha} = \begin{pmatrix}0&\bar\sigma^{\alpha}\\\sigma^\alpha&0\end{pmatrix}
\]
where $\sigma^\alpha\equiv(I,\sigma^i)$ is the Pauli 4-vector and $\bar\sigma^\alpha\equiv(I,-\sigma^i)$. The Weyl representation allows us to extract from the Dirac algebra $\{\gamma^\alpha,\gamma^\beta\}=2\eta^{\alpha\beta}$ the left-handed two-component algebra
\begin{align}
\sigma^\alpha_{\ AA'}\bar\sigma^{\beta A'B}+\sigma^\beta_{\ AA'}\bar\sigma^{\alpha A'B}=2\eta^{\alpha\beta}{\delta}_A^{\ B}\label{dalg}
\end{align}
where $\delta_{A}^{\ B}$ is the Kronecker delta.  If we use the convention in \cite{Bailin} whereby the primed index for $\bar{\sigma}^\alpha$ is a row index and unprimed is a column index, and the opposite for $\sigma^\alpha$, we can explicitly identify both $\bar\sigma^{0 A'A} = \delta^{A'A}$ and $\sigma^0_{\ AA'}=\delta_{AA'}$ as the $2\times2$ identity matrix, and the spatial parts $\bar{\sigma}^{iA'A}$ and $\sigma^i_{\ AA'}$ as the usual Pauli matrices.

The Pauli 4-vector plays a special role because it is invariant under Lorentz transformations on all indices, that is  \cite{DHM2010},
\begin{equation}
\Lambda^\alpha_{\ \beta}\Lambda^{A}_{\ B}\bar\Lambda^{A'}_{\ B'} \bar{\sigma}^{\beta B'B}=\bar{\sigma}^{\alpha A'A}\label{LIsigma}
\end{equation}
where $\Lambda^\alpha_{\ \beta}$ is an arbitrary spin-$1$ Lorentz transformation and $\Lambda^{A}_{\ B}$ is the corresponding spin-$\half$ Lorentz transformation.

The generators of the group are constructed as $S^{\alpha\beta} = \frac\ii4 [\gamma^{\alpha},\gamma^{\beta}]$ for the 4-component formalism or in spinor notation
\begin{equation}
{L^{\alpha\beta}}_A^{\ B}=\frac\ii4 \left(\sigma^\alpha_{\ AA'}\bar\sigma^{\beta A'B}-\sigma^\beta_{\ AA'}\bar\sigma^{\alpha A'B}\right)\label{gen}
\end{equation}
for the left-handed 2-spinor.

Note that the Pauli 4-vector is not referred to as an operator in this formalism. An operator for spinors $\psi_B$ instead carries an index structure $A_{A}^{\ B}$. Wherever possible we will keep indices implicit and use the standard notation $\hat A$ for operators. Using this notation the components of the generators can be written as
\begin{align*}
\hat L^{0j}=&\frac\ii2 \hat\sigma^{j}, \qquad
\hat L^{ij}=\frac12\eps^{ij}_{\ \ k}\hat\sigma^k
\end{align*}
where $\hat\sigma^{i}$ are the standard Pauli operators.\footnote{While having the same components as the Pauli matrices, note the distinction between the {\it operator} $\hat\sigma^{i}$ which maps a spinor $\psi_{A}\to\phi_{A}$ whereas the object $\bar{\sigma}^{iA'A}$ would map $\psi_{A}\to\phi^{A'}$.} The $\hat L^{0j}$ components generate boosts and the $\hat L^{ij}$ components generate rotations.

\subsection{Lorentz invariant measurement formalism\label{sec-unitarity}\label{sec-measurement}}

Up to this point we have discussed spinor notation and the spin-$\half$ representation of the Lorentz group. This structure on its own does not provide a quantum formalism. To achieve a quantum mechanical description we must also introduce an inner product to promote the spinor space $W$ to a Hilbert space $\m H$, and then construct a formalism to extract predictions according to the rules of quantum mechanics. This formalism differs to that used in the Wigner representation since the two component irreducible representation of $SL(2,\mathbb C)$ constitutes a {\it non-unitary} spin-$\half$ representation of the Lorentz group. However, a notion of unitarity can be recovered by introducing a suitable inner product and not insisting on using representations. This has the effect of modifying the form of Hermitian operators, as well as causing the action of the Lorentz group on qubits to no longer be a representation. The latter is an immaterial effect, whereas the former requires a reformulation of Hermitian operators. We will here simply state the results of the formalism derived in \cite{PTW11}.

\subsubsection{The quantum state and inner product\label{newinnerproduct}\label{secfermionQS}}\label{appIP}

In order to promote the spinor space $W$ to a Hilbert space $\m H$ an inner product is required. In spinor notation a sesquilinear inner product requires a spinorial object with index structure $I^{A'A}$. The appropriate object is given by $I_{u}^{A'A}\equiv u_\alpha\bar{\sigma}^{\alpha A'A}$ where $u_{\alpha}$ is the 4-velocity of the particle carrying the spin. For a Wigner state this would be represented by the momentum $p$ in $\ket{p,\psi}$. The inner product between states represented by the two spinors $\psi_A^{1}$ and  $\psi_A^{2}$ is
\begin{eqnarray}
\bk{\psi^{1}}{\psi^{2}}=I_{u}^{A'A}\bar{\psi}^{1}_{A'}\psi^{2}_A=u_\alpha\bar{\sigma}^{\alpha A'A}\bar{\psi}^{1}_{A'}\psi^{2}_A
\label{IP}
\end{eqnarray}
where the connection between Dirac bra-ket notation and spinor notation is identified as
\begin{equation}
|\psi\rangle\sim\psi_A\qquad\langle\psi|\sim I_{u}^{A'A}\bar{\psi}_{A'}.\label{braket}
\end{equation}
The inner product \eqref{IP} is Lorentz invariant. This follows immediately from the fact that all indices have been contracted, and that $\bar\sigma^{\alpha A'A}$ is invariant under Lorentz transformation, \eqref{LIsigma}. This inner product emerges by taking the WKB limit of the Dirac field. In this limit not only do we obtain a well defined inner product on the spinor space and a classical trajectory with velocity $u^{\alpha}$ for a localised wavepacket but in addition a conserved probability current $j^{\alpha}$ that ultimately allows us to apply a quantum mechanical interpretation to the state $\psi_{A}$. This allows us to promote $W$ to a Hilbert space. Given that $I_{u}^{A'A}$ depends on the particle's 4-velocity or equivalently 4-momentum, the corresponding Hilbert space is labelled with momentum $p$:  $\m H_{p}$. With this inner product we have a finite dimensional unitary formalism describing the transformation of spin  $\psi_{A}$ under arbitrary Lorentz transformations. Note that we no longer strictly have a representation of the Lorentz group since an arbitrary Lorentz transformation $\Lambda$ will correspond to a map between distinct Hilbert spaces $\Lambda: \m H_{p}\to \m H_{\Lambda p}$, rather than within a single space.\footnote{We recall that a representation of a group consists of the set of linear operators acting on a single vector space.} By not insisting on using representations we have managed to sidestep Wigner's theorem that any faithful unitary representation of the Lorentz group must be infinite dimensional \cite{Wigner}.

\subsubsection{Hermitian operators \label{sec-hermitian}}

Now that we have a well-defined inner product, we can consider the general form of Hermitian operators, and derive the mathematical form of observables. To do this we start with the standard Hermitian property for an operator $\hat A$:
\begin{align*}
0 &= \bk{\chi}{A\psi} - \bk{A\chi}{\psi}\\
   &= I_{u}^{A'A}\bar\chi_{A'} A_A^{\ B}\psi_B - I_{u}^{A'A} \bar A_{A'}^{\ B'}\bar\chi_{B'} \psi_A.
\end{align*}
For this to hold for all $\chi_{A}$ and $\psi_{A}$ we must have that $I_{u}^{A'B}A_B^{\ A} = I_{u}^{B'A} \bar A_{B'}^{\ A'}$. Making use of \eqref{dalg} and \eqref{gen}, and using the self dual property $\hat L^{\alpha\beta}=\half \ii\eps^{\alpha\beta\gamma\delta}\hat L_{\gamma\delta}$ \cite[Eqn.\,2.74]{DHM2010} to introduce the {\it Pauli-Lubanski vector} $\hat W^\alpha(p):=\half\eps^{\alpha\beta\gamma\delta}p_\beta \hat L_{\gamma\delta} = \ii p_\beta  \hat L^{\alpha\beta}$, one can then show that a Hermitian operator $\hat A$ must be of the form
\begin{equation}
\hat A= 2\ii n_\alpha u_\beta  \hat L^{\alpha\beta}+n_\alpha u_\beta \eta^{\alpha\beta} \hat {\m I}=n_\alpha \left(-\frac{2\hat W^\alpha(p)}{m}+u^{\alpha}\hat{\m I}\right)\label{fermionoperator}
\end{equation}
where each operator is identified by a Lorentz 4-vector $n_\alpha$ of real coefficients. Every Hermitian operator has a real eigenvalue spectrum, and its eigenstates $\ket{\psi^\pm}$ are orthogonal with respect to the inner product $I_u^{A'A}$.

We are now interested in spin observables formed from Hermitian operators \eqref{fermionoperator}. A spin observable evaluated in the particle's rest frame should reduce to the non-relativistic expression $n_i\sigma^i$ where $n^i$ is the normalised spin measurement direction. This implies that $n^\alpha$ in \eqref{fermionoperator} is orthogonal to $u^\alpha$, i.e.\;we have the condition
\be
u_\alpha n^\alpha=0.\label{eq-orthogcond}
\ee
The magnitude of $n^\alpha$ only rescales the eigenvalues of the observable and without loss of generality we can normalise it so that it is spacelike with $n^2=-1$. The Lorentz invariant expectation value of an observable $\hat A$ for a spinor $\psi_A$ is then given by
\be
\langle\psi|\hat A|\psi\rangle=-I_u^{A'A}\bar{\psi}_{A'} n_\alpha \frac{2W^{\alpha B}_{\ A} }m \psi_B=-n_\alpha\bar\sigma^{\alpha A'A}\bar\psi_{A'}\psi_A. \label{spacetimemeas}
\ee
This expression is covariant and has been written in the Weyl representation where we have made use of the relationship $\frac 2m I_u^{A'B}W^{\alpha A}_{\ B}=\bar\sigma^{\alpha A'A}-u^\alpha u_\beta\bar\sigma^{\beta A'A}$.\footnote{For those familiar with the Wigner representation, the expectation value $\bra{k,\psi}\mathbf n\!\cdot\mathbf{\hat\sigma}\ket{k,\psi}$ in the rest frame is written as $-2\bra{p,\psi} n_\alpha \hat W^\alpha(p)/m\ket{p,\psi}$ in a boosted frame, with $\hat W^\alpha(p)/m=\half  L(p)^\alpha_{\ i}\hat\sigma^i$  \cite{Bogoliubov,Ternotworol,Czachor}, where $L(p)$ is the boost relating the frames.} Lorentz invariance follows immediately because all spinor and spacetime indices have been contracted, and all objects transform covariantly.

\section{Intuitive derivation of the Stern-Gerlach observable}\label{SGOP}

The task now is to determine the correct spin observable for a Stern-Gerlach measurement in which the Stern-Gerlach apparatus and particle carrying the spin have a relativistic relative velocity. Given the measurement formalism and formation of Lorentz invariant expectation values outlined in the previous section, the problem of determining  a `relativistic Stern-Gerlach spin operator' is reduced to simply determining how $n_{\alpha}$ is related to the direction of the Stern-Gerlach apparatus. This analysis will follow \cite{PTW11}.

\subsection{The non-relativistic Stern-Gerlach experiment\label{sec-nonrelSG}}

Before we consider the relativistic case it is helpful to first review a standard non-relativistic Stern-Gerlach spin measurement represented by the arbitrary non-relativistic Hermitian observable $n_{i}\hat\sigma^{i}$. In this case a particle is passed though an inhomogeneous magnetic field. This causes the wavepacket to separate into two packets of orthogonal spin, after which a position measurement records the outcome. In the rest frame of the particle, the fermion is exposed to a magnetic field $B^\SG_i = |B^\SG| b^\SG_{i}$ for a short period of time, which is the magnetic field as measured specifically in the Stern-Gerlach rest frame, denoted by $\text{SG}$. The direction $b^\SG_{i}$ of the magnetic field defines the quantization direction of the spin, and the {\it gradient} of the magnetic field $\nabla_{i}|B^\SG|$ determines the rate and direction along which the wavepacket splits into eigenstates of $b^\SG_{i}\hat\sigma^{i}$ \cite{PeresQM}. Therefore in the non-relativistic case the measurement direction $n_{i}$ is simply the direction of the magnetic field in the Stern-Gerlach rest frame, $b^\SG_i$.

\subsection{The relativistic Stern-Gerlach experiment}

We now turn to the relativistic scenario. In this case the qubit is now moving through the Stern-Gerlach apparatus with relativistic velocity. Viewed in the rest frame of the particle, denoted $\text{RF}$, the measurement process is indistinguishable from the non-relativistic one described in \S\ref{sec-nonrelSG}. However, in this frame the fermion will experience a {\it transformed} magnetic field $B_{i}^{\RF} = |B^{\RF}|b_{i}^{\RF}$. The measurement direction is now given by $n^i=b^i_\RF$, giving a spin observable of $b_i^\RF\sigma^i$.\footnote{Similarly it is the gradient of the rest frame magnetic field $\nabla_{i}|B^{\RF}|$ which now determines the rate and direction along which the wavepacket splits.} This spin observable is written in the specific frame of the particle rest frame, and due to the transformation properties of the magnetic field, the relationship between $b^\RF_i$ and the orientation of the Stern-Gerlach apparatus is nontrivial. The goal is now to arrive at a covariant expression of the spin observable in terms of the Stern-Gerlach direction $b_{\alpha}^\SG$ and the 4-velocities of the Stern-Gerlach apparatus and the particle.

In order to do this we assume that the electromagnetic field generated consists of purely a magnetic field in the rest frame of the Stern-Gerlach device,
\begin{equation}
F_{\alpha\beta} = -\epsilon_{\alpha\beta\gamma\delta}v^{\gamma}B^{\delta}_\SG\overset{**}{=}\begin{pmatrix}0&0\\0&B_{ij}\end{pmatrix}\label{SGemfield}
\end{equation}
where $v^{\gamma}$ is the 4-velocity of the Stern-Gerlach device and $B^{\delta}_\SG$ is the magnetic field 4-vector of the Stern-Gerlach device. The double star `$**$' of the right hand side indicates that it has been evaluated explicitly in the frame where $v^{\alpha} \overset{**}{=}(1,0,0,0)$, i.e.\;the Stern-Gerlach rest frame in which the Stern-Gerlach  magnetic 4-vector is given by $B^\delta_\SG\overset{**}{=}(0,B^i_\SG)$.

We can now determine the 4-vector $B_{\RF}^\alpha$ defined by $B^\alpha_\RF\overset*=(0,B^i_\RF)$, where `$*$' indicates evaluation in the particle rest frame, $u^\alpha\overset*=(1,0,0,0)$. The covariant expression for $B^\alpha_\RF$ is given by $B_{\RF}^\alpha\equiv-\frac12\epsilon^{\alpha\beta\gamma\delta}u_\beta F_{\gamma\delta}$, which when inserted into \eqref{SGemfield} yields
\begin{equation}
B_{\RF}^\alpha=\frac12\epsilon^{\alpha\beta\gamma\delta}u_\beta \epsilon_{\gamma\delta \kappa\lambda}v^\kappa B_\SG^\lambda=B_\SG^\alpha(v\cdot u)-v^\alpha(B_\SG\cdot u)\label{BRF}
\end{equation}
with $\epsilon^{\alpha\beta\gamma\delta} \epsilon_{\gamma\delta \kappa\lambda}=-2(\delta^\alpha_\kappa\delta^\beta_\lambda-\delta^\alpha_\lambda\delta^\beta_\kappa)$ \cite[p.87]{MTW}, and the notation $a\cdot b\equiv a_\alpha b^\alpha$ indicating a 4-vector scalar product. Considering a spin measurement using this magnetic field, the four-vector $n^\alpha$ in \eqref{spacetimemeas} is now the normalized direction of the rest frame magnetic field:
\begin{equation}
n^\alpha(m,u,v)\equiv b_{\RF}^\alpha = \frac{B_{\RF}^\alpha}{|B_{\RF}|}
\label{SGn}
\end{equation}
where $|B_{\RF}|:=\sqrt{-B_{\RF}^\alpha B_{\RF}^\beta\eta_{\alpha\beta}}$, and from \eqref{BRF} we have $b_{\RF}\cdot u=0$, in agreement with \eqref{eq-orthogcond}. By Eqn.\eqref{fermionoperator} the relativistic spin operator is therefore given by
\begin{equation}
\m S_{A}^{\ B}:=-b^\RF_\alpha \frac{{2W^\alpha}_A^{\ B}}m\label{SG_spin_op},
\end{equation}
and the expectation value of the corresponding measurement is calculated using \eqref{spacetimemeas}. Expectation values \eqref{spacetimemeas} are invariant under simultaneous Lorentz transformation of both particle and apparatus, and thus with only the relative velocity between the apparatus and qubit and the spatial orientation of the Stern-Gerlach apparatus, we can calculate the expectation values corresponding to a relativistic Stern-Gerlach spin measurement.

\section{WKB analysis of a Stern-Gerlach measurement}\label{WKB}

In section \ref{SGOP} we argued that in the particle rest frame the measurement process is indistinguishable to a non-relativistic one \cite{PeresQM,Ballentine}. From this analysis we saw that it is the direction of the magnetic field, as seen in the particle rest frame, that determines which spin measurement is being carried out. That derivation used standard arguments based on classical relativity and nonrelativistic quantum mechanics  (see also \cite{PTW11,Pablo1111}).

In this section we provide a more fundamental analysis of the relativistic Stern-Gerlach measurement process, using the minimally coupled Dirac equation in the WKB limit. Our starting point is an equivalent two component formulation of the Dirac equation called the {\it Van der Waerden} equation. The two component field is then expanded, without loss of generality, as a superposition in the eigenbasis of the spin operator \eqref{SG_spin_op}. Using the linearity of the Van der Waerden equation, we can analyse each component of the superposition separately, and identify the classical trajectories of each component. We will use this to show that an inhomogeneous magnetic field results in the splitting of a localised wave-packet with arbitrary spin into two components of orthogonal spin. This analysis singles out \eqref{SG_spin_op} as the relevant spin operator for a relativistic Stern-Gerlach spin measurement.

\subsection{The WKB equations}

A qubit physically realized by the spin of a massive fermion is described by the Dirac field, therefore our starting point will be the Dirac equation minimally coupled to the electromagnetic field
\begin{eqnarray}
\ii\gamma^\alpha D_\alpha\Psi=\ii\gamma^\alpha \left(\partial_\alpha-\ii eA_{\alpha}\right)\Psi=m\Psi\label{curveddiraceq}
\end{eqnarray}
where we define the $U(1)$ covariant derivative as $D_\alpha\equiv\partial_\alpha-\ii eA_\alpha$. $\Psi$ is the Dirac field, $\gamma^{\alpha}$ are the Dirac $\gamma$-matrices and $A_{\alpha}$ is the electromagnetic four potential. Strictly speaking it is the positive frequency solutions of the Dirac equation that describe a one particle state. Furthermore we assume that the strength of the electric field is insufficient to cause particle creation. We refer the reader to \cite{PTW11} for a discussion of spin of a massive fermion as a realisation of a relativistic qubit.

We proceed with the WKB approximation by putting the Dirac equation into a second order form. In the Weyl representation of the Dirac matrices, the field splits into $\Psi=(\phi_A,\chi^{A'})$ \cite{PeskinSchroeder}. The objects $\phi_A$ and $\chi^{A'}$ are each two-component Weyl-spinor fields constituting left- and right- handed spinor representations of $SL(2,\mathbb C)$. In this representation the Dirac equation splits into two separate equations
\begin{subequations}
\begin{align}
\ii\bar{\sigma}^{\alpha A'A}D_\alpha\phi_A&=m\chi^{A'}\label{left}\\
\ii\sigma^{\alpha}_{\ AA'}D_\alpha\chi^{A'}&=m\phi_A\label{right}.
\end{align}
\end{subequations}
Solving for $\chi^{A'}$ in equation \eqref{left}, inserting the result into \eqref{right}, and rearranging yields a second order equation called the {\it Van der Waerden equation} \cite{SakuraiAQM}
\begin{equation}
\eta^{\alpha\beta} D_\alpha D_\beta\phi_A- eF_{\alpha\beta}L^{\alpha\beta\ B}_{\ \ A}\phi_B+m^2\phi_A = 0\label{vdweqn}
\end{equation}
where $F_{\alpha\beta}=\partial_\alpha A_\beta-\partial_\beta A_\alpha$ is the electromagnetic tensor and we have used that $\hat{L}^{\alpha\beta} = \frac \ii2\sigma^{[\alpha}\bar{\sigma}^{\beta]}$ and $\hat\eta^{\alpha\beta}=\sigma^{\{\alpha}\bar\sigma^{\beta\}}$.

The next step is to consider the Van der Waerden equation in the high frequency WKB  limit \cite{PTW11}. In this limit we see that the fermion travels along classical trajectories. We will assume that the field is sufficiently localised for the purposes of the Stern-Gerlach measurement.\footnote{We refer the reader to \cite{PTW11} for further details on localisation.} The goal then is to show that the wavepacket is split by the Stern-Gerlach magnetic field into two packets of spin corresponding exactly to the eigenstates of the spin operator \eqref{SG_spin_op}.

Traditional treatments of the WKB approximation begin with an ansatz for the spinor field of the form
\[
\phi_{A}(x) =  \varphi_{A}(x)\Ee^{\ii\theta(x)/\eps}
\]
where $\eps$ is to be thought of as a `dummy' parameter whose only role is to identify the different orders in an expansion. This ansatz is substituted into the Van der Waerden equation. One can then expand in the limit $\eps\rightarrow0$; the high frequency limit. Mathematically this corresponds to splitting the field into a rapidly varying phase $\theta$ and a slowly varying envelope $\varphi_{A}$. The phase determines a field of wavevectors $k_{\alpha}\equiv \partial_{\alpha}\theta -e A_{\alpha}$ which in this limit define integral curves along which the envelope is transported.

In the case of a Stern-Gerlach spin measurement we know the initial wave packet will be split into two wavepackets of orthogonal spin with different wavevectors $k^{\pm}_{\alpha}\equiv \partial_{\alpha}\theta^{\pm} -e A_{\alpha}$. Therefore we must slightly modify the WKB ansatz to
\begin{equation}
\phi_{A} = a \varphi^{+}_{A}\Ee^{\ii\theta^{+}/\eps}+b \varphi^{-}_{A}\Ee^{\ii\theta^{-}/\eps}\label{wkb_sol}
\end{equation}
where we have, without loss of generality, expanded the spinor field in the eigenbasis of $\m S_{A}^{\ B}$ \eqref{SG_spin_op} with components $a,b \in\mathbb C$ defined so that $|a|^{2}+|b|^{2} = 1$. The integral curves are determined by the phases  $\theta^{\pm}$ which correspond to the spin eigenstates $\varphi_{A}^{\pm}$. We will see that these components are deflected in two different directions by the inhomogeneous magnetic field.

Using the linearity of the Van der Waerden equation we can analyze each component $\phi^{\pm}_{A} = \varphi^{\pm}_{A}\Ee^{\ii\theta^{\pm}/\eps}$ separately. Substituting $\phi^{\pm}_{A}$ into \eqref{vdweqn} yields
\begin{multline}
\left(\eta^{\alpha\beta}\partial_\alpha \partial_\beta\varphi^{\pm}_A- eF_{\alpha\beta}L^{\alpha\beta\ B}_{\ \ A}\varphi^{\pm}_B+\frac{\ii}{\eps}(2k_{\pm}^\alpha\partial_\alpha\varphi^{\pm}_A+\varphi^{\pm}_A\partial_\alpha k_{\pm}^\alpha) \right.\\\left.
-\frac{1}{\eps^2}k^{\pm}_\alpha k_{\pm}^\alpha\varphi^{\pm}_A+m^2\varphi^{\pm}_A\right)\Ee^{\ii\theta^{\pm}/\eps} = 0 \label{vdwwkb}
\end{multline}
where $k^{\pm}_{\alpha}\equiv \partial_{\alpha}\theta^{\pm} -e A_{\alpha}$. It is customary to treat the mass term as $\eps^{-2}$ and we shall do so here. In the $\eps\rightarrow0$ limit, which corresponds to large momentum, we notice that the electromagnetic field term $F_{\alpha\beta}$ has a negligible influence in \eqref{vdwwkb}. Thus in order for the fermion to `feel' the presence of the electromagnetic field, we will need to treat $F_{\alpha\beta}$ as a $1/\eps$ term.

The WKB approximation proceeds by separating the orders of $\eps$. For our purposes we neglect the lowest order terms and thus obtain the following set of equations
\begin{align}
\frac 1\eps\left(2k_{\pm}^\alpha\partial_\alpha\varphi^{\pm}_A+\varphi^{\pm}_A\partial_\alpha k_{\pm}^\alpha+\ii eF_{\alpha\beta}L^{\alpha\beta\ B}_{\ \ A}\varphi^{\pm}_B\right)&=0, \label{wkb1}\\
\frac 1{\eps^{2}}\left(k^{\pm}_\alpha k_{\pm}^\alpha-m^2\right)\varphi^{\pm}_A &=0. \label{wkb2}
\end{align}
The first equation \eqref{wkb1} will describe the evolution of the spin state of $\varphi_A^\pm$ along a trajectory \cite{PTW11}. The second equation \eqref{wkb2} determines the trajectories along which the fermion is transported. As it is, the spin does not couple to the magnetic field, implying that the trajectories cannot be spin-dependent, and thus no spin-dependent deflection of packets can occur. However, if we treat the gradient of the magnetic field as a $\eps^{-2}$ term, we can include such a term in \eqref{wkb2}:
\begin{align}
\left(k^{\pm}_\alpha k_{\pm}^\alpha-m^2\right)\varphi^{\pm}_A - \eps eF_{\alpha\beta}L^{\alpha\beta\ B}_{\ \ A}\varphi^{\pm}_B&=0. \label{wkb3}
\end{align}
To zeroth order in $\eps$, Eqn.\eqref{wkb3} is still the standard dispersion relation, and implies that $k_{\pm}^{\alpha}$ is timelike. However, upon taking the gradient of \eqref{wkb3}, the second term becomes relevant. It is in this way that the trajectories become spin-dependent, producing the separation of the wavepacket that occurs in a Stern-Gerlach measurement.

\subsection{Determining the spin-dependent trajectories}

It is from the integral curves of $\frac{\di x^\alpha_{\pm}}{\di\tau}=u_{\pm}^\alpha(x)\equiv k_{\pm}^\alpha(x)/m$ that we can read off the deflection of the trajectories, where $u^{\alpha}_{+} = u^{\alpha}_{-}$ prior to entering the magnetic field. In order to deduce the implications of \eqref{wkb3} we first multiply it by $I_{u_\pm}^{A'A}\bar\varphi^{\pm}_{A'}$, obtaining
\begin{align}
\left(k^{\pm}_\alpha k_{\pm}^\alpha-m^2\right)|\varphi^{\pm}|^{2} - \eps eF_{\alpha\beta}I_{u_\pm}^{A'A}L^{\alpha\beta\ B}_{\ \ A}\varphi^{\pm}_B\bar\varphi^{\pm}_{A'}&=0. \label{wkb4}
\end{align}
Next, we decompose the operator $F_{\alpha\beta}{\hat L^{\alpha\beta}}$ into the electric field $\hat{E}^{\RF}$ and magnetic field $\hat{B}^{\RF}$ operators as measured in the rest frame defined by the initial 4-velocity $u^{\pm}_{\alpha}$. This is given by
\be
\begin{split}
F_{\alpha\beta}\hat L^{\alpha\beta}&=\hat E^\RF+\hat B^\RF\\
& \equiv 2u^{\pm}_{\gamma}u_{\pm}^\alpha F_{\alpha\beta}\hat L^{\gamma\beta}+F_{\alpha\beta}{h_{\pm}}_{\ \gamma}^{\alpha}{h_{\pm}}_{\ \delta}^{\beta} \hat L^{\gamma\delta}\label{electromagnetic_term}
\end{split}
\ee
where ${h_{\pm}}_{\ \gamma}^{\alpha} \equiv \delta_{\ \gamma}^{\alpha}-u^{\alpha}_{\pm}u^{\pm}_{\gamma}$ is the spacetime projector onto the space orthogonal to the 4-velocity $u_{\pm}^\alpha$. The first term is anti-Hermitian with respect to the inner product $I_{u_\pm}^{A'A}$, whereas the second term is Hermitian.

First consider the magnetic field term $\hat{B}^{\RF}$. Using the self-dual property $\hat L^{\alpha\beta}=\half \ii\eps^{\alpha\beta\gamma\delta}\hat L_{\gamma\delta}$, and substituting the specific form \eqref{SGemfield} of the electromagnetic field of the Stern-Gerlach apparatus, the expression can be rearranged to give

\begin{equation}
\begin{split}
\hat B^\RF=& F_{\alpha\beta}{h_{\pm}}_{\ \gamma}^{\alpha}{h_{\pm}}_{\ \delta}^{\beta}
\hat L^{\gamma\delta}\\
 =& -2\ii u_{\alpha}B^{\RF}_{\beta}\hat L^{\alpha\beta}\equiv \m |B^\RF|\hat{\m S}\label{magneticoperator}
\end{split}
 \end{equation}
where $|B_{\RF}|$  is the magnitude of the magnetic field as measured in the rest frame of $u^{\pm}_{\alpha}$ \eqref{BRF}. We see that the magnetic field operator is in fact the relativistic spin operator \eqref{SG_spin_op} derived in the previous section multiplied by the field strength. Given that $\varphi^\pm_A$ are defined as eigenstates of this operator, \eqref{wkb_sol}, we therefore have that
\be
\bra{\psi^{\pm}}\hat{B}^{\RF}\ket{\psi^{\pm}}=\pm |B^{\RF}|\label{magneticterm}
 \ee
where the quantum state is identified as $\ket{\psi^{\pm}} \sim\varphi^{\pm}_{A}/|\varphi^{\pm}|$ with $|\varphi^\pm|^2\equiv\bar\varphi^\pm_{A'}I_{u_\pm}^{A'A}\varphi_A^\pm$.

Let us now proceed to show that the expectation values of $\hat E_\RF$ in \eqref{electromagnetic_term} with $\ket{\psi^\pm}$ are zero. Firstly, we define the projector $\hat\Pi^\pm_{B_\RF}\equiv\half(\hat{\m I}\pm\frac1{|B_\RF|^2}\hat B_\RF)$, so we have
\[
\bra{\psi^\pm}\hat E_\RF\ket{\psi^\pm}=\Tr[\hat E_\RF\kb{\psi^\pm}{\psi^\pm}]=\Tr[\hat E_\RF\hat\Pi^\pm_{B_\RF}] =\Tr[\hat E_\RF \half(\hat{\m I}\pm\frac1{|B_\RF|^2}\hat B_\RF)].
\]
Using the Lorentz invariance of the expectation values, we can evaluate the expectation values in the particle rest frame where the operators take on the form $\hat E^\RF\overset*=E_i^\RF\hat\sigma^i$ and $\hat B^\RF\overset*=B_i^\RF\hat\sigma^i$:
\[
\bra{\psi^\pm}\hat E_\RF\ket{\psi^\pm}\overset*=\pm\frac1{|B^\RF|^2}E^\RF_iB_\RF^i.
\]
The electric field vector is given by  $E_\RF^i\overset*=F^{0i}=\eps^{ijk}v_jB_k^\SG$. By Eqn.\eqref{BRF}, $B_\SG^i$ is a linear combination of $B_\RF^i$ and $v^i$, and so we have $E_\RF^i B^\RF_i=0$. Therefore
\begin{equation}
\bra{\psi^{\pm}}\hat{E}^{\RF}\ket{\psi^{\pm}}= 0.\label{electricterm}
 \end{equation}

Now substituting \eqref{magneticterm} and \eqref{electricterm} into \eqref{wkb4}, and taking the gradient of the resulting equation, we obtain
\begin{align}
0 & =  \partial_{\alpha}(k^{\pm}_\beta k_{\pm}^\beta)\pm\partial_{\alpha}(|B_{\RF}|)\nonumber\\
& =  2k_{\pm}^\beta\partial_{\beta}k^{\pm}_\alpha + 2ek^{\beta}_{\pm}F_{\beta\alpha}\pm\partial_{\alpha}(|B_{\RF}|)\label{trajectory}
\end{align}
where we have used that $\partial_{\alpha}(|B_{\RF}|)\sim 1/\eps^{2}$. We see that $\frac{\di x_{\pm}^\alpha}{\di\tau}=u_{\pm}^{\alpha}=k_{\pm}^{\alpha}/m$ must satisfy
\begin{eqnarray}
m\frac{\di^2x_{\pm}^\alpha}{\di\tau^2}+e\frac{\di x_{\pm}^\beta}{\di\tau} F_\beta^{\ \alpha}\pm \frac1m\partial^{\alpha}(|B_{\RF}|)=0\label{FWvelocity}
\end{eqnarray}
where $\frac{\di^2x_{\pm}^\alpha}{\di\tau^2}=\frac{\di x_{\pm}^\beta}{\di\tau}\partial_\beta u^\alpha_{\pm}$ is the 4-acceleration and $u_{\pm}^\alpha u^{\pm}_\alpha=1$. The first two terms of \eqref{FWvelocity} are simply the classical Lorentz force law, but in addition to this we have a deflection induced by the non-zero magnetic field gradient $\pm\partial_{\alpha}(|B_{\RF}|)/m$ whose sign depends on whether the spin is parallel or anti-parallel to the magnetic field $B^{\RF}_{\alpha}$.

The implications of \eqref{FWvelocity} are as follows: prior to measurement we have that  $u_{+}^{\alpha}=u_-^{\alpha}$. The qubit is then exposed to a strongly inhomogeneous electromagnetic field $F_{\alpha\beta}$ for a short period of time. This impulse-like interaction alters the velocity of the respective packets. For an ideal measurement this interaction is short enough that negligible precession of the spin, governed by \eqref{wkb1}, will occur during this splitting. The end result is the deflection of the $\psi_{A}^+$ component of spin with amplitude $|a|^{2}$ in the direction of the gradient of the magnetic field, and the deflection of the $\psi_A^-$ component with amplitude $|b|^2$ in the opposite direction. A position measurement will then produce the outcome `$+$' with probability $|a|^2$, and `$-$' with $|b|^2$. Thus we conclude that the operator corresponding to relativistic Stern-Gerlach measurement is given by \eqref{SG_spin_op}.

\section{Conclusion and discussion}\label{conclusion}

This paper provided two distinct ways of identifying the spin observable corresponding to a Stern-Gerlach measurement of a massive fermion where the relative velocity of the particle and Stern-Gerlach apparatus is relativistic. The first approach followed an intuitive argument based on the transformation properties of the electromagnetic field. The second approach was a first-principles approach, starting with the Dirac equation minimally coupled to the electromagnetic field. Using this equation we showed that in the WKB limit the `$+$' spin eigenstate of \eqref{SG_spin_op} is deflected `up' and the `$-$' spin eigenstate is deflected `down'. We therefore concluded that in the relativistic regime the appropriate spin-operator for a Stern-Gerlach measurement is
\be
 \hat{\m S}(v,p,b^\SG)= - \frac{b^\SG_{\alpha}(v\cdot p)-v_{\alpha}(b^\SG\cdot p)}{\sqrt{(v\cdot p)^2-(b^\SG\cdot p)^2}}\frac{2\hat W^\alpha(p)}m.
\label{concl-SGspinop}
\ee
Notably the spin operator \eqref{concl-SGspinop} is momentum-dependent, so that, if the momentum is unknown, it is not possible to determine the expectation value. This has the following implications: Firstly, tracing over momentum of a state written in a tensor product basis of spin and momentum has been used in relativistic quantum information theory to extract the reduced spin density matrix \cite{PeresScudoTerno02,BartlettTerno05,Friis-relent,GingrichAdami02,PeresTerno03b}. However, we can see that due to the momentum dependence of the observable \eqref{concl-SGspinop}, this reduced spin density matrix is not useful for extracting statistics of a relativistic Stern-Gerlach measurement. The usefulness of the reduced spin density matrix is further limited by the fact  that it has no Lorentz covariant transformation properties \cite{PeresScudoTerno02,PeresTerno03b,TernoIRQI,PeresTerno04}.

Secondly, in quantum tomography of spin, one collects measurement data for various measurement directions. One then uses this data to solve for the quantum state. Non-relativistically, the relationship between the quantum state and measurement data is momentum independent, and it is enough to choose three linearly independent directions in order to reconstruct the quantum state. However, in the relativistic Stern-Gerlach case, the experimental data are related to momentum dependent theoretical expectation values of the quantum state, determined by \eqref{concl-SGspinop}. Thus, if momentum is unknown, three linearly independent directions will not suffice. We leave it as an open question as to what minimal set of measurements is required to reconstruct the state in this relativistic case.

As a final point about the specific form of \eqref{concl-SGspinop}, we note that in the literature there exist several alternative operators that are used as observables for relativistic spin measurements. Two notable operators that have been proposed in the relativistic quantum information community are $\hat{\m S}'\propto a_{i}\hat W^i$ \cite{Czachor,Friis-relent} and $\hat{\m S}'' \propto a_{i }(\hat W^i-\hat W^0p^i/(p^0+m))$ \cite{Ternotworol,Landulfo,Caban1,Caban2,LCY04,KimSon,Lee03,Moradi10,CRW09,RS09}, where $a_i$ is a parameter determining which measurement is carried out. Although these proposals are Hermitian they are mathematically distinct from \eqref{concl-SGspinop} and it can be shown that they lead to quantitatively distinct predictions. An intuitive reason for this can be found in \cite{Pablo1111} where the authors show that the directions extracted from the $\hat{\m S}'$ operator do not transform in the same way as a magnetic field. It can also be shown in a similar analysis that the directions extracted from $\hat{\m S}''$ do not transform like a magnetic field. As a result these proposals cannot be considered to represent a Stern-Gerlach measurement. The question that therefore must be addressed is whether there exists a physical implementation for either of these proposals. However, measurements making use of a coupling of the spin to the electromagnetic field will not yield these spin operators.

\section*{Acknowledgments}

We would like to thank Stephen Bartlett, Daniel Terno, Aharon Brodutch, and Nikolai Friis for stimulating and helpful discussions. This research was supported by the Perimeter Institute-Australia Foundations (PIAF) program, and the Australian Research Council grant DP0880860.

\bibliography{RQI_RSO,RQI}

\begin{thebibliography}{10}

\bibitem{PTW11}
M.~C. Palmer, M.~Takahashi, and H.~F. Westman, ``Localized qubits in curved
  spacetimes,'' {\em Annals of Physics}, vol.~327, no.~4, pp.~1078--1131, 2012.

\bibitem{Pablo1111}
P.~L. Saldanha and V.~Vedral, ``Physical interpretation of the wigner rotations
  and its implications for relativistic quantum information,'' {\em New Journal
  of Physics}, vol.~14, no.~2, p.~023041, 2012.

\bibitem{Czachor}
M.~Czachor, ``Einstein-podolsky-rosen-bohm experiment with relativistic massive
  particles,'' {\em Phys. Rev. A}, vol.~55, pp.~72 -- 77, Jan 1997.

\bibitem{Ternotworol}
D.~R. Terno, ``Two roles of relativistic spin operators,'' {\em Phys. Rev. A},
  vol.~67, p.~014102, Jan 2003.

\bibitem{PeresTerno04}
A.~Peres and D.~R. Terno, ``Quantum information and relativity theory,'' {\em
  Rev. Mod. Phys.}, vol.~76, pp.~93 -- 123, Jan 2004.

\bibitem{PeresScudoTerno02}
A.~Peres, P.~F. Scudo, and D.~R. Terno, ``Quantum entropy and special
  relativity,'' {\em Phys. Rev. Lett.}, vol.~88, p.~230402, May 2002.

\bibitem{BartlettTerno05}
S.~D. Bartlett and D.~R. Terno, ``Relativistically invariant quantum
  information,'' {\em Phys. Rev. A}, vol.~71, p.~012302, Jan 2005.

\bibitem{Caban1}
P.~Caban and J.~Rembieli\ifmmode~\acute{n}\else \'{n}\fi{}ski,
  ``Lorentz-covariant reduced spin density matrix and
  einstein-podolsky-rosen-bohm correlations,'' {\em Phys. Rev. A}, vol.~72,
  p.~012103, Jul 2005.

\bibitem{Caban2}
P.~Caban and J.~Rembieli\ifmmode~\acute{n}\else \'{n}\fi{}ski,
  ``Einstein-podolsky-rosen correlations of dirac particles - quantum field
  theory approach,'' {\em Phys. Rev. A}, vol.~74, p.~042103, 2006.

\bibitem{Friis-relent}
N.~Friis, R.~A. Bertlmann, M.~Huber, and B.~C. Hiesmayr, ``Relativistic
  entanglement of two massive particles,'' {\em Phys. Rev. A}, vol.~81,
  p.~042114, Apr 2010.

\bibitem{Landulfo}
A.~G.~S. Landulfo and G.~E.~A. Matsas, ``Influence of detector motion in bell
  inequalities with entangled fermions,'' {\em Phys. Rev. A}, vol.~79,
  p.~044103, Apr 2009.

\bibitem{Alsing}
P.~M. Alsing, J.~C. Evans, and K.~K. Nandi, ``The phase of a quantum mechanical
  particle in curved spacetime,'' {\em Gen. Relativ. Gravit.}, vol.~33,
  pp.~1459 -- 1487, Oct 2000.

\bibitem{TerashimaUeda03}
H.~Terashima and M.~Ueda, ``Einstein-podolsky-rosen correlation in a
  gravitational field,'' {\em Phys. Rev. A}, vol.~69, p.~032113, Mar 2004.

\bibitem{Louko}
J.~Louko and A.~Satz, ``How often does the unruh--dewitt detector click?
  regularization by a spatial profile,'' {\em Classical and Quantum Gravity},
  vol.~23, no.~22, p.~6321, 2006.

\bibitem{Fuentes}
I.~Fuentes, R.~B. Mann, E.~Mart\'in-Mart\'inez, and S.~Moradi, ``Entanglement
  of dirac fields in an expanding spacetime,'' {\em Phys. Rev. D}, vol.~82,
  p.~045030, Aug 2010.

\bibitem{Brodutch11}
A.~Brodutch and D.~R. Terno, ``Polarization rotation, reference frames, and
  mach's principle,'' {\em Phys. Rev. D}, vol.~84, p.~121501, Dec 2011.

\bibitem{BDT11}
A.~Brodutch, T.~F. Demarie, and D.~R. Terno, ``Photon polarization and
  geometric phase in general relativity,'' {\em Phys. Rev. D}, vol.~84,
  p.~104043, Nov 2011.

\bibitem{FoldyWouthuysen}
L.~L. Foldy and S.~A. Wouthuysen, ``On the dirac theory of spin 1/2 particles
  and its non-relativistic limit,'' {\em Phys. Rev.}, vol.~78, pp.~29--36, Apr
  1950.

\bibitem{HehlNi}
F.~W. Hehl and W.-T. Ni, ``Inertial effects of a dirac particle,'' {\em Phys.
  Rev. D}, vol.~42, pp.~2045--2048, Sep 1990.

\bibitem{Ryder98}
L.~H. {Ryder}, ``{Relativistic treatment of inertial spin effects},'' {\em
  Journal of Physics A Mathematical General}, vol.~31, pp.~2465--2469, Mar.
  1998.

\bibitem{Ryder99}
L.~H. {Ryder}, ``{Relativistic Spin Operator for Dirac Particles},'' {\em
  General Relativity and Gravitation}, vol.~31, pp.~775--780, May 1999.

\bibitem{Mashhoon95}
B.~{Mashhoon}, ``{On the coupling of intrinsic spin with the rotation of the
  earth},'' {\em Physics Letters A}, vol.~198, pp.~9--13, Feb. 1995.

\bibitem{Weinberg}
S.~Weinberg, {\em The Quantum Theory of Fields, Vol. 1: Foundations}.
\newblock Cambridge University Press, 1~ed., June 1995.

\bibitem{Wald}
R.~M. Wald, {\em General Relativity}.
\newblock University Of Chicago Press, 1~ed., June 1984.

\bibitem{Bailin}
D.~Bailin and A.~Love, {\em {Supersymmetric Gauge Field Theory and String
  Theory}}.
\newblock Taylor \& Francis, 1~ed., Oct. 1994.

\bibitem{Penrose}
R.~Penrose and W.~Rindler, {\em Spinors and {Space-Time}: Volume 1,
  {Two-Spinor} Calculus and Relativistic Fields (Cambridge Monographs on
  Mathematical Physics)}.
\newblock {Cambridge University Press}, Feb. 1987.

\bibitem{DHM2010}
H.~K. Dreiner, H.~E. Haber, and S.~P. Martin, ``Two-component spinor techniques
  and feynman rules for quantum field theory and supersymmetry,'' {\em Phys.
  Rept.}, vol.~494, pp.~1--196, 2010.

\bibitem{Wigner}
E.~Wigner, ``On unitary representations of the inhomogeneous lorentz group,''
  {\em Ann. Math.}, vol.~40, no.~1, pp.~149 -- 204, 1939.

\bibitem{Bogoliubov}
N.~Bogoliubov and D.~Shirkov, {\em Introduction to axiomatic quantum field
  theory}.
\newblock Wiley-Interscience, 1980.

\bibitem{PeresQM}
A.~Peres, {\em Quantum Theory: Concepts and Methods (Fundamental Theories of
  Physics)}.
\newblock Springer, Oct. 1993.

\bibitem{MTW}
C.~W. Misner, K.~S. Thorne, and J.~A. Wheeler, {\em Gravitation (Physics
  Series)}.
\newblock W. H. Freeman, 1~ed., Sept. 1973.

\bibitem{Ballentine}
L.~E. Ballentine, {\em {Quantum Mechanics: A Modern Development}}.
\newblock World Scientific Publishing Company.

\bibitem{PeskinSchroeder}
M.~E. Peskin and D.~V. Schroeder, {\em An Introduction To Quantum Field Theory
  (Frontiers in Physics)}.
\newblock Westview Press, Oct. 1995.

\bibitem{SakuraiAQM}
J.~J. Sakurai, {\em Advanced Quantum Mechanics}.
\newblock Addison-Wesley, Reading, Massachusetts, 1980.

\bibitem{GingrichAdami02}
R.~M. Gingrich and C.~Adami, ``Quantum entanglement of moving bodies,'' {\em
  Phys. Rev. Lett.}, vol.~89, p.~270402, Dec 2002.

\bibitem{PeresTerno03b}
A.~Peres and D.~R. Terno, ``Quantum information and special relativity,'' {\em
  Int. J. Quant. Info.}, vol.~1, 225, 2003.

\bibitem{TernoIRQI}
D.~R. Terno, {\em Quantum Information Processing: From Theory to Experiment,
  edited D.G. Angelakis et al.}, pp.~61--86.
\newblock IOP Press, 2006.

\bibitem{LCY04}
D.~Lee and E.~Chang-Young, ``Quantum entanglement under lorentz boost,'' {\em
  New Journal of Physics}, vol.~6, no.~1, p.~67, 2004.

\bibitem{KimSon}
W.~T. Kim and E.~J. Son, ``Lorentz-invariant bell's inequality,'' {\em Phys.
  Rev. A}, vol.~71, p.~014102, Jan 2005.

\bibitem{Lee03}
D.~Ahn, H.-j. Lee, Y.~H. Moon, and S.~W. Hwang, ``Relativistic entanglement and
  bell's inequality,'' {\em Phys. Rev. A}, vol.~67, p.~012103, Jan 2003.

\bibitem{Moradi10}
S.~Moradi and M.~Aghaee, ``Frame independent nonlocality for three qubit
  state,'' {\em Int. J. Theor. Phys.}, vol.~49, pp.~615--620.

\bibitem{CRW09}
P.~Caban, J.~Rembieli\ifmmode~\acute{n}\else \'{n}\fi{}ski, and
  M.~W\l{}odarczyk, ``Strange behavior of the relativistic
  einstein-podolsky-rosen correlations,'' {\em Phys. Rev. A}, vol.~79,
  p.~014102, Jan 2009.

\bibitem{RS09}
J.~Rembielin\'ski and K.~A. Smolin\'ski, ``Quantum preferred frame: Does it
  really exist?,'' {\em Europhysics Letters}, vol.~88, no.~1, p.~10005, 2009.

\end{thebibliography}

\end{document}